\documentstyle[epsf,aps,multicol]{revtex}
\def\frac#1#2{{#1 \over #2}}
\def\half{{\frac 12}}
\def\third{{\frac 13}}
\def\fourth{{\frac 14}}

\def\sixth{{\frac 16}}
\def\beq{\begin{equation}}
\def\eeq{\end{equation}}
\def\etal{{\it et al.\ }}

\def\deg{^{\circ}}

\def\Tna{$T_{NA}$}

\def\m{\mathrm{m}}
\def\K{\mathrm{K}}

\def\mK{\mathrm{mK}}

\def\T{\mathrm{T}}

\def\cm{\mathrm{cm}}
\def\mm{\mathrm{mm}}

\def\j2rel{\mathrel{\mkern-4mu}}

\def\gtorder{\mathrel{\raise.4ex\hbox{$>$}\mkern-14mu
             \lower0.6ex\hbox{$\sim$}}}
\def\ltorder{\mathrel{\raise.4ex\hbox{$<$}\mkern-14mu
             \lower0.6ex\hbox{$\sim$}}}

\begin{document}

\title{Two Experimental Tests of the Halperin-Lubensky-Ma Effect at the Nematic--Smectic-A Phase Transition}

\author{Anand Yethiraj and John Bechhoefer}
\address{Department of Physics, Simon Fraser University, Burnaby, British Columbia, V5A 1S6, Canada}


\maketitle

\begin{abstract}
We have conducted two quantitative tests of predictions based on
the Halperin-Lubensky-Ma (HLM) theory of fluctuation-induced
first-order phase transitions.  First, we explore the effect of an
external magnetic field on the nematic--smectic-A (NA) transition
in a liquid crystal.  Second, we examine the dependence of the
first-order discontinuity as a function of mixture concentration
in pure 8CB and three 8CB-10CB mixtures.  We find the first
quantitative evidence for deviations from the HLM
theory.\end{abstract}

\begin{multicols}{2}
One of the most important advances made in our understanding of
continuous phase transitions has been the modification of critical
exponents, due to thermal fluctuations, from the values predicted
by mean-field theory \cite{pfeuty}. But thermal fluctuations have
another effect, one that is less-well-understood theoretically and
studied only to a limited extent experimentally: when two order
parameters (or an order parameter and a gauge field) are
simultaneously present and interact with each other, the
fluctuations of one may drive the phase transition of the other
first order.  In high-energy physics, for example, an analogous
situation occurs in the Higgs mechanism \cite{pfeuty}.  In
condensed-matter physics, over two decades ago, Halperin, Lubensky
and Ma (HLM) \cite{HLM} predicted that this could occur in two
settings: the normal-superconducting phase transition in type-1
superconductors and the nematic--smectic-A (NA) transition in
liquid crystals. At the NA transition
\cite{dGP,garland2,anisimov3,cladis2}, the first fluctuating field
is the nematic order parameter $Q_{ij}= S\left(3n_{i}n_{j} -
\delta_{ij}\right)/2$, while the other order parameter is the
two-component smectic-A order parameter $\psi$. In principle, a
complete theory must account for thermal fluctuations in the
nematic order parameter magnitude $S$, the nematic director
$\hat{n}$, and the smectic-A order parameter $\psi$.  The
de~Gennes-McMillan mechanism \cite{dGP} takes into account $S$
fluctuations but is mean field in $\hat{n}$ and $\psi$.  The HLM
theory, in addition, considers $\hat{n}$ fluctuations but is mean
field in $\psi$. Deviations from the de~Gennes-McMillan mechanism
have been quantitatively established by Anisimov \etal
\cite{anisimov2} and Cladis \etal \cite{cladis1}.  Although their
conclusions are consistent with the HLM mechanism, the experiments
are at their resolution limits and the full implications of the
HLM theory remain to be tested.

In this paper, we conduct the first two quantitative tests of the
HLM theory, both leading to significant discrepancies from HLM.
(1) Mukhopadhyay \etal \cite{ranjan} have made a detailed
prediction for the external-field dependence of the HLM effect; we
look for this field effect experimentally in the cyanobiphenyl
liquid crystal 8CB, as well as in mixtures of 8CB and 10CB.  (2)
The first-order discontinuity implied by the HLM theory (which can
be characterized by the number $t_{0} = (T_{NA} - T^{*})/T^{*}$,
where $T_{NA}$ is the transition temperature and $T^{*}$ is the
spinodal point) should also be sensitive to the mixture
concentration $x$ in the 8CB-10CB system.  Here, we make a
quantitative study of $t_{0}$ vs. $x$.

In order to achieve the resolution required for these measurements, we
designed a high-resolution real-space optical technique (``Intensity Fluctuation
Microscopy'') which is at least an order of magnitude more sensitive
than previous techniques.  The experimental set-up is briefly
described next (see Ref.~\cite{yethiraj,yethirajthesis} for more
details).

The liquid crystal is sandwiched between two glass cubes, both treated
for unidirectional planar anchoring.  The spacing between the cubes
(typically $d = 30\,\mu\m$) is adjusted interferometrically, and
thickness gradients across the entire cell are controlled to be
$\ltorder 0.5 \,\mu\m$.  The cubes are placed in a gradient hot-stage
specifically designed for this experiment to fit an in-house
transverse-bore electromagnet.  The top and bottom of the sample cell
are independently temperature controlled at temperatures that straddle
the NA transition temperature.

We use a crossed-polarized digital CCD microscopy set-up to
measure intensity fluctuations as a finely resolved
function of temperature.  These intensity fluctuations are caused by
director fluctuations which are a soft mode in the nematic phase but
are suppressed in the smectic-A phase.
Subtracting two successive, uncorrelated ``instantaneous'' images
(a $1.8\,\mm \times 1.4\,\mm$ area of the sample being imaged onto a
$1317 \times 1024$ pixel, 12-bit digital CCD camera)
gives us a difference image
\beq
\Delta(x,y) = I_{1}(x,y) - I_{2}(x,y),
\eeq
which reflects static intensity fluctuations.  The histogram of the
intensity differences in the nematic and smectic-A phase is well fit
by a Gaussian probability distribution.  The primary contribution to
the noise is photoelectronic shot-noise, which also has a Gaussian
probability distribution.  The {\it normalized variance}, $\zeta$ has the
shot-noise variance subtracted and is normalized with respect to the
transmitted light intensity:
\beq
\zeta \,=\, \frac{\hat{\zeta}
-\hat{\zeta}_{SN}}{(\langle\overline{I}\rangle - I_{offset})^{2}},
\eeq
where $\hat{\zeta}$ is the measured {\it raw} variance,
$\hat{\zeta}_{SN}$ is the shot-noise variance calibrated in a blank
field, $\langle\,\rangle$ refers to a spatial average over pixels,
$\langle\overline{I}\rangle$ is the average intensity over both
images and all pixels, and $I_{offset}$ is an built-in intensity
offset in the CCD camera.

The variance $\zeta$ decreases as the NA transition temperature,
$T_{NA}$, is approached from above.
In the smectic-A phase, it is much smaller; in fact, $\zeta$ is not
measurably higher than the zero (shot-noise) level.
Since the applied temperature gradient straddles the NA transition,
the resulting difference image contains a hot nematic part and a
cooler smectic-A part separated by a flat boundary.  We divide our image into strips parallel to
the NA boundary, which is typically flat over the field-of-view.
By calibration, we translate the spatial dependence of $\zeta$ into a
temperature dependence.  Each image that contains an NA interface is {\it
self-calibrating} in that over small times, the gradient remains
constant while the overall temperature fluctuates slowly.
The r.m.s. fluctuations in temperature over long times ($\approx 1$
hour) was roughly
$0.15$ mK and was less than $0.05$ mK over the duration of a
measurement.

The experimental technique described above was first used to study the
critical behavior close to $T_{NA}$.  Pure 8CB is a material that
exhibits a NA transition that is indistinguishable from second order
when studied by traditional high-resolution calorimetry.  Dynamical
measurements by Cladis \etal \cite{cladis1} have suggested that this
transition is
first order.  We confirm here (see also \cite{yethiraj}) that
this transition is unambiguously first order.
\begin{figure}
\par\columnwidth 20.5pc
\hsize\columnwidth\global\linewidth\columnwidth
\displaywidth\columnwidth
\epsfxsize=3.0truein
\centerline{\epsfbox{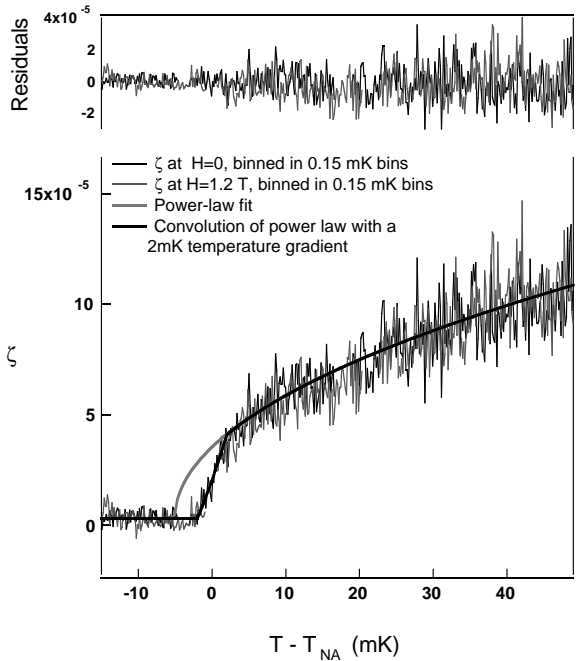}}
\caption{\protect\small Pure 8CB, $d = 30\,\mu\m$, $0.59 \pm 0.02 \K/\cm$.
Here, $t_{0} = \frac{5.0 \pm 0.5
\,\mK.}{307\,\K} = 1.6 \pm 0.1\times10^{-5}$.  Results are shown with
and without an external applied field.}
\label{fig:8CBMF2b}
\end{figure}
The fluctuation-vs.-temperature profile
for 8CB is plotted in Fig.~\ref{fig:8CBMF2b} (black data curve).
There are three distinct sections to
the graph:
\begin{enumerate}
\item Smectic: Here, $\zeta = 3.1 \pm 3.0 \times 10^{-6}$, and
the smectic fluctuation level is flat and indistinguishable from the
shot-noise background.
\item Nematic: The data here fit a power law of the form
\beq
w_{0} + w_{1}(T - w_{2})^{w_{3}}
\label{eq:powlaw}
\eeq
with $w_{0}= 3.1 \pm 3.0 \times 10^{-6}$, $w_{1} = 1.43 \pm 0.04 \times 10^{-5}$,
$w_{2} = 5.0\pm 0.5 \,\mK$, and $w_{3} = 0.50\pm 0.05$.
The fit (shown in Fig.~\ref{fig:8CBMF2b}) was done over the largest
temperature range that still kept the residuals (top curve in
Fig.~\ref{fig:8CBMF2b}) flat.  $w_{0}$ was held to the smectic level,
while
$w_{3}$ is the exponent of the power law.  $w_{2}$, the divergence
point of the power law, is identified with the spinodal temperature
$T^{*}$.  However, it is a phantom
divergence point, as the actual phase transition is located at the
interface, i.e., at $T - T_{NA} = 0\,\mK$ \cite{exponent}
%
 \item Interface: Ideally, the interface on Fig.~\ref{fig:8CBMF2b} would
 be a vertical line.  However, a temperature gradient normal to the sample 
plane smears out the interface.  Varying temperature moves the position of 
the interface in the field-of-view, but the magnitude of the smearing is 
stable to within the statistical errors in the data.  In practice, we fit the 
entire data to a convolution of the power
law form in Eq.~\ref{eq:powlaw} and a linear temperature gradient.  This fit 
differs from the simple power law fit (also shown) only in the interfacial 
region.  The position of the interface is unambiguously determined by the fit 
and does NOT coincide with the zero point.  The top curve in 
Fig.~\ref{fig:8CBMF2b} shows the residuals for such a fit.
\end{enumerate}

We tested for the effect of sample thickness $d$ on the measurements.  We
find no systematic dependence on sample thickness for $d >
7.5\,\mu\m$.  For smaller $d$, we find a marked {\it increase}
in $t_{0}$, which we ascribe to the effect of a small residual
rotation ($\approx 0.2 \deg$) between the anchoring direction of the top and bottom
plates.  The twist created by this rotation is strongest for small
$d$.  The implications of this will be examined in a
longer article \cite{yethiraj2}.

Having established the experimental technique, we first study the
critical effects of suppressing director fluctuations with an
external magnetic field applied along the easy axis of the nematic.
Magnetic-field effects in the nematic phase have been studied
in the last two decades \cite{durand,poggi1}.  Scale fields to change
Landau parameters at the NA transition are large; the mean-field
effects of such large fields have been studied by Lelidis and Durand
\cite{lelidis4}.  Recently, Mukhopadhyay \etal \cite{ranjan} have shown that based
on the HLM mechanism, one expects modest fields to suppress the
director fluctuations that drive a mean-field second-order transition first
order.  They predict a non-analytic dependence of $t_{0}$ on the
field $H$ and a critical field, $H_{c}$, above which the transition is
second order.  In particular, they predict
\beq
H_{c}= H_{0}\,t_{0}^{1/2},
\eeq
where $H_{0}$ is a scale field that depends on known material
parameters.  For 8CB, $H_{0}\approx 3500\,\T$ and from
Fig.~\ref{fig:8CBMF2b}, $t_{0}= 1.6 \times 10^{-5}$: this implies
a critical field of $H_{c}\approx 10\T$.

The magnetic field effect at the transition was probed with a field
that could be varied from $0\,\T$ to $1.5\,\T$.  Contrary to initial
expectations, no suppression of the fluctuations was seen within
$50\,\mK$ of the transition point.  Two data curves are shown in
Fig.~\ref{fig:8CBMF2b}.  The data curves (black representing zero field,
and grey representing measurements at a field of $1.2\,\T$)
practically fall on top of each other.  Likewise, we see no shift in
$t_{0}$.

What do the magnetic field studies on a $0\,$--$1.5\,T$ field tell us
about the critical field of $\approx 10\,\T$ predicted via the HLM
scenario?
A weaker effect
implies a larger critical field.

To establish an lower bound for $H_{c}$, we look at the small-$|H|$
limit of the HLM prediction,
\beq
t_{0}(H)/t_{0} \approx 1 - m\,H/H_{c},
\eeq
where $m = (2.17 \pm 0.01)$ is the slope of the linear suppression
predicted in Ref.~\cite{ranjan}.
In our case, our
confidence level for $t_{0}$ is roughly $10\%$; i.e., the largest
suppression of $t_{0}$ that we would {\it not observe} is
$t_{0}(H)/t_{0} = 0.9$.
This puts a lower bound on the value of $H_{c}$.
The maximum field used is
$H = H_{max} = 1.5 \,\T$.
Thus, for 8CB, we find
\begin{eqnarray}
H_{c}&\geq& \frac{m\,H}{1 - t_{0}(H)/t_{0}}\nonumber \\
     &=& \frac{(2.17 \pm 0.01)\,1.5 \,\T}{0.9}\nonumber \\
     &\approx& 33 \,\T.
\end{eqnarray}
Thus, the lower bound for the critical field in our measurements is
three times the predicted value.

We also measured the effect of a magnetic field in two mixtures,
0.18 and 0.41 mole fraction 10CB in 8CB.  In
neither of these samples do we observe a suppression of fluctuations
close to $T_{NA}$.

One might worry about the validity of a null result.  We do,
however, see the expected {\it non-critical} suppression of the
fluctuations.  We find that this depression falls off quadratically with increasing field:
\beq
\zeta = \zeta_{0} - g_{H} H^{2}.
\eeq
This non-critical reduction in the field-effect on approaching $T_{NA}$
reflects quite clearly the fact that the (twist and bend) elastic constants
are getting stiffer as one approaches \Tna.  In fact, since the magnetic
coherence length
is $\xi_{H} = {({\frac{K}{\xi}})}^{\half}{\frac{1}{H}}$, one reasonably expects
that the field dependence is always a function of $\xi_{H}$; i.e., the
coefficient must obey
$g_{H} \propto {\frac{1}{K}}$, where $K$ is a diverging elastic
constant.  Fitting to the value of $\bar{\nu} = 0.5$ that is
consistent with other experiments and with our earlier fit in zero
field, we find that the data clearly agree with this (see also
\cite{yethiraj2}.)

Our second experimental test of the HLM effect was the dependence of
$t_{0}$ on the concentration $x$ of 10CB in 8CB, between $x = 0$ and $x =
0.44$.  This dependence has been studied previously by adiabatic
calorimetry (Marynissen \etal \cite{maryn1}).  Anisimov \etal
\cite{anisimov2} determined the Landau tricritical point (the LTP is
the tricritical point predicted by the de~Gennes-McMillan theory) to
be at $x^{\star}=0.41$.  The resolution in the adiabatic calorimetry
is, however, limited to $x > 0.3$.\cite{cladisnote} To test the HLM
theory unambiguously, we extended the range of mixture concentration
to $x = 0$ (pure 8CB).  The non-zero latent heat for $x < x^{\star}
\approx 0.41$ implies that their data are inconsistent with the
de~Gennes-McMillan theory but consistent with the HLM form, which
contains a small negative cubic term (with a constant cubic
coefficient B) that keeps the latent heat non-zero.  In HLM, the effective
mean-field Landau free energy is
\beq
f= \half A |\psi|^{2} - \third B |\psi|^{3} + \fourth C |\psi|^{4} +
\sixth E |\psi|^{6}
\label{eq:Lhlm}
\eeq
In Ref.~\cite{anisimov2}, the condition for coexistence ($f = 0$ and
$dF/d\psi = 0$) was used to
calculate the entropy change $\Delta S = L/T$, in terms of the three
Landau parameters $A$, $B$, and $C$.  Close to the transition, $A
\equiv \alpha t$, where $t$ is the reduced temperature, and $\alpha$ is
a constant.  The variation of the quartic coefficient near the LTP is
modeled by $C = C_{0}(x - x^{\star})$.
To make connection with our measurements of $t_{0}$, we
calculate
$t_{0}\equiv A_{NA}/\alpha$,
 where $A_{NA}$, the quadratic coefficient at the transition, is a
function of $B$, $C$, and $E$.
\begin{figure}
\par\columnwidth 20.5pc
\hsize\columnwidth\global\linewidth\columnwidth
\displaywidth\columnwidth \epsfxsize=3.0truein
\centerline{\epsfbox{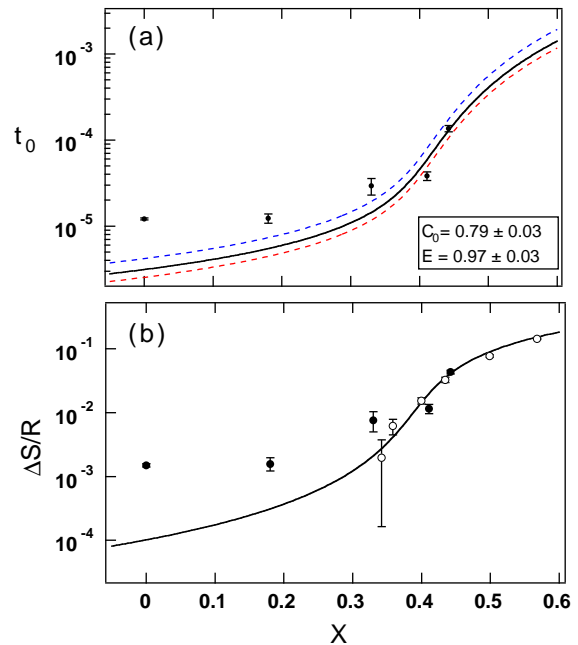}}
\caption{\protect\small (a) Fit of
$t_{0}$ data to the Anisimov parameters, with $C_{0}$ as a fit
parameter.  Top and bottom dashed curves show fits to $t_{0}$ for
$\bar{\nu}=0.6$ and $\bar{\nu}=0.4$, respectively.  (The data
points for these exponent values are not shown.)  (b) Comparison
of latent heat data (Ref.~16) to $t_{0}$ data (this work).  Open
circles: data taken from Ref.~16; filled circles: $t_0$ data
converted to equivalent latent heat.}
\label{fig:t0lh}
\end{figure}
The fit parameters from the analysis of Anisimov \etal
\cite{anisimov2} are
\begin{eqnarray}
\beta &\equiv& -3/8 (\alpha C_{0}/E) = 0.993\nonumber\\
&&\Delta S^{\star}/R = 0.0261
\end{eqnarray}
While the entropy change per mole $\Delta S/R$ depends only on $B$, $\alpha
C_{0}$ and $E$, the discontinuity $t_{0}$ depends on all four parameters
separately: here, we use the two Anisimov parameters and
fit $C_{0}$ and $E$.

Each data point in Fig.~\ref{fig:t0lh}(a) is obtained from the
power-law fit described earlier.  The exponent of the power law is
fixed at $\bar{\nu} = 0.5$.  The error bars on a free fit are $\approx
0.1$.  In Fig.~\ref{fig:t0lh}(a) we plot the data for a fitted
exponent of $\bar{\nu} = 0.5$ and the HLM fit.  Also shown, for
perspective, are HLM fits for $t_{0}$ data with exponents $\bar{\nu} =
0.4$ and $\bar{\nu} = 0.6$.  The values of $C_{0}$ in the three fits
are $-0.74 \pm 0.03$, $-0.79 \pm 0.03$, and $-0.88 \pm 0.03$,
respectively.  (The best-fit value of $E=0.97 \pm 0.03$ for $\bar{\nu}
= 0.5$ is used for all three fits.) In Fig.~\ref{fig:t0lh}(b), we
compare the latent-heat data of Marynissen \etal \cite{maryn1} with
that calculated from our $t_{0}$ data (using the Anisimov parameters).
The discontinuity gets larger as we approach the Landau tricritical
point.  In particular, we find that the data near the LTP agree with
the HLM prediction for reasonable values of the quartic coefficient
$C_{0}$.  However, for lower concentrations of 10CB in 8CB, there is a
clear deviation from the HLM prediction, which {\it underestimates}
the latent heat in this region of the phase diagram \cite{fitnote}.

Two main results arise out of this study, and both constrain theories
in the weakly type-1 region of the NA transition.  The first is the
absence of a magnetic field effect on the critical behavior in pure
8CB and in mixtures.  It was predicted \cite{ranjan} that an
external field would suppress nematic director fluctuations and drive
the transition back to second order.  Experimentally, we see a
non-critical suppression of the fluctuations but observe no change in
the critical behavior.  In particular, the strength of the
first-order transition, measured by the dimensionless number $t_{0}$,
is unchanged.  We thereby establish a lower bound of $30\,\T$ on the
critical magnetic field, three times higher than the predicted value.
We conclude that the neglect of smectic fluctuations
by HLM is not
valid in this region of the phase diagram.

Second, the variation of $t_{0}$ in mixtures has yielded the
surprising result that $t_{0}$ on the weakly first-order side of the
LTP is larger than the HLM predicted value, and not smaller as one
might naively expect of a crossover to second-order XY behavior.
This might be related to the observed local
maximum in critical exponents seen at the 3DXY to tricritical
crossover \cite{garland2}. This
result puts a strong constraint on any theory of the
tricritical point.

Both results in this work provide, for the first time, quantitative
evidence for deviation from the implications of both the
de~Gennes-McMillan theory and the HLM theory.
 Our data
show not only that the discontinuity is {\it larger} than that
predicted by HLM theory but also that the field dependence is much
weaker than predicted.  These results provide a testing ground for
any future theory that goes beyond HLM; the logical next step
would be to account for the effect of smectic fluctuations. Future
experimental work will focus on the effect of larger fields.

This work was supported by NSERC (Canada).
We acknowledge useful discussions with R. Mukhopadhyay and P. Cladis,
and thank B.~Heinrich for generous access to his magnet.


%
%
%


\begin{thebibliography}{10}

\bibitem{pfeuty}
P. Pfeuty and G. Toulouse, {\em Introduction to the Renormalization Group and
  to Critical Phenomena}, 1st  ed. (Wiley-Interscience, Chichester, 1978).  Cf. p. 153.

\bibitem{HLM}
B.~I. Halperin, T.~C. Lubensky, and S.~K. Ma, Phys.\ Rev.\ Lett.\ {\bf 32},
  292  (1974).

%
\bibitem{dGP}
P.~G. de~Gennes and J. Prost, {\em The Physics of Liquid Crystals}, 2nd  ed.
  (Clarendon Press, Oxford, 1993).

\bibitem{garland2}
C.~W. Garland and G. Nounesis,  Phys.\ Rev.\ E {\bf 49}, 2964 (1994).

\bibitem{anisimov3}
M.~A. Anisimov,  Mol.\ Cryst.\ Liq.\ Cryst.\ {\bf 162A}, 1 (1988)

\bibitem{cladis2}
P.~E. Cladis, in {\em Physical Properties of Liquid Crystals}, {D.
Demus {\it et~al.}}, ed., (Wiley-VCH, 1999), \ pp.\ 277.

\bibitem{anisimov2}
M.~A. Anisimov, P.~E. Cladis, E.~E. Gorodetskii, D.~A. Huse, V.~E.
Podneks, V.~G. Taratuta, W. van~Sarloos, and V.~P. Voronov, Phys.\ Rev.\ A {\bf 41},  6749  (1990).

\bibitem{cladis1}
P.~E. Cladis, W. van~Sarloos, D.~A. Huse, J.~S. Patel, J.~W. Goodby,
and P.~L. Finn, Phys.\ Rev.\ Lett.\ {\bf 62},  1764  (1989).

\bibitem{ranjan}
R. Mukhopadhyay, A. Yethiraj, and J. Bechhoefer, Phys.\ Rev.\ Lett.\
(submitted) (1999).

\bibitem{yethiraj}
A. Yethiraj and J. Bechhoefer, Mol. Cryst. Liq. Cryst. {\bf 304},  301  (1997).

\bibitem{yethirajthesis}
A. Yethiraj, Ph.D. thesis, Simon Fraser University, 1999.

\bibitem{exponent}
$\bar{\nu}$ is a phenomenological fit parameter.  The real form
\cite{yethiraj2} is more complicated, and the fitted value
$\bar{\nu}=0.5$ is merely used to extrapolate data to set $T^{\star}$.

\bibitem{yethiraj2}
A. Yethiraj, R. Mukhopadhyay and J. Bechhoefer, in preparation.

\bibitem{durand}
J.~L.~Martinand and G.~Durand, Sol.\ St.\ Commun.\ {\bf 10,}
  815 (1972).

\bibitem{poggi1}
Y. Poggi and J.~C. Filippini, Phys.\ Rev.\ Lett.\ {\bf 39,}
  150 (1977).

\bibitem{lelidis4}
I. Lelidis and G. Durand, Phys.\ Rev.\ Lett.\ {\bf 73},  672  (1994).


\bibitem{maryn1}
H. Marynissen, J. Thoen, and W. van Dael, Mol. Cryst. Liq. Cryst. {\bf 124},
  195  (1985).

\bibitem{cladisnote}
The dynamical calorimetry of Ref.~\cite{cladis1} extended all the way
to pure 8CB; however, the large error bars there limited quantitative
comparison with theory.

\bibitem{fitnote}
One possible explanation for this disagreement is that the linear
variation $C = C_{0}(x - x^{\star})$ might not be valid; allowing for a
nonlinear (quadratic or cubic) correction term in the fit yields a
coefficient that is larger than $C_{0}$, which is unphysical; this fit
is also qualitatively {\it worse} (see Ref.~\cite{yethiraj2}).




%
%
%


%


\end{thebibliography}

\end{multicols}
\end{document}